\documentclass[pra,showpacs,amsfonts,amsmath,twocolumn,superscriptaddress,a4paper]{revtex4}
\usepackage{array,amsmath,amsfonts,amssymb,dsfont} \usepackage{natbib}
\usepackage[T1]{fontenc} \usepackage{ae} \usepackage{color}
\usepackage{graphicx}

\begin{document}

\title{Manipulating Rydberg atoms close to surfaces at cryogenic temperatures}

\author{T. Thiele}
\email{tthiele@phys.ethz.ch}
\affiliation{Department of Physics, ETH Zurich, CH-8093 Zurich, Switzerland}

\author{S.~Filipp}
\email{filipp@phys.ethz.ch}
\affiliation{Department of Physics, ETH Zurich, CH-8093 Zurich, Switzerland}

\author{J.~A. Agner}
\affiliation{Laboratorium f\"ur Physikalische Chemie, ETH Z\"urich, CH-8093 Z\"urich, Switzerland}

\author{H. Schmutz}
\affiliation{Laboratorium f\"ur Physikalische Chemie, ETH Z\"urich, CH-8093 Z\"urich, Switzerland}

\author{J. Deiglmayr}
\affiliation{Laboratorium f\"ur Physikalische Chemie, ETH Z\"urich, CH-8093 Z\"urich, Switzerland}

\author{M. Stammeier}
\affiliation{Department of Physics, ETH Zurich, CH-8093 Zurich, Switzerland}

\author{P. Allmendinger}
\affiliation{Laboratorium f\"ur Physikalische Chemie, ETH Z\"urich, CH-8093 Z\"urich, Switzerland}

\author{F. Merkt}
\affiliation{Laboratorium f\"ur Physikalische Chemie, ETH Z\"urich, CH-8093 Z\"urich, Switzerland}

\author{A.~Wallraff} \affiliation{Department of Physics, ETH Zurich, CH-8093 Zurich, Switzerland}


\renewcommand{\i}{{\mathrm i}} \def\1{\mathchoice{\rm 1\mskip-4.2mu
l}{\rm 1\mskip-4.2mu l}{\rm 1\mskip-4.6mu l}{\rm 1\mskip-5.2mu l}}
\newcommand{\ket}[1]{|#1\rangle} \newcommand{\bra}[1]{\langle #1|}
\newcommand{\braket}[2]{\langle #1|#2\rangle}
\newcommand{\ketbra}[2]{|#1\rangle\langle#2|}
\newcommand{\opelem}[3]{\langle #1|#2|#3\rangle}
\newcommand{\projection}[1]{|#1\rangle\langle#1|}
\newcommand{\scalar}[1]{\langle #1|#1\rangle}
\newcommand{\op}[1]{\hat{#1}} \newcommand{\vect}[1]{\boldsymbol{#1}}
\newcommand{\id}{\text{id}}

	\begin{abstract}
		Helium atoms in Rydberg states have been manipulated coherently with microwave radiation pulses near a gold surface and near a superconducting NbTiN surface at a temperature of $3~\text{K}$. The experiments were carried out with a skimmed supersonic beam of metastable $(1\text{s})^1(2\text{s})^1\, {}^1\text{S}_0$ helium atoms excited with laser radiation to $n\text{p}$ Rydberg levels with principal quantum number $n$ between $30$ and $40$. The separation between the cold surface and the center of the collimated beam is adjustable down to $250~\mu\text{m}$. Short-lived $n\text{p}$ Rydberg levels were coherently transferred to the long-lived $n\text{s}$ state to avoid radiative decay of the Rydberg atoms between the photoexcitation region and the region above the cold surfaces. Further coherent manipulation of the $n\text{s}$ Rydberg levels with pulsed microwave radiation above the surfaces enabled measurements of stray electric fields and allowed us to study the decoherence of the atomic ensemble. Adsorption of residual gas onto the surfaces and the resulting slow build-up of stray fields was minimized by controlling the temperature of the surface and monitoring the partial pressures of H$_2$O, N$_2$, O$_2$ and CO$_2$ in the experimental chamber during the cool-down. Compensation of the stray electric fields to levels below $100~\text{mV}/\text{cm}$ was achieved over a region of $6~\text{mm}$ along the beam-propagation direction which, for the $1770~\text{m}/\text{s}$ beam velocity, implies the possibility to preserve the coherence of the atomic sample for several microseconds above the cold surfaces.	
	\end{abstract}
	
	\maketitle
	
	\section{Introduction} 
The interest in coherent manipulation of Rydberg atoms has grown in recent years in several fields of physics. For example, coherent collective Rydberg excitations have been demonstrated in ultracold atom clouds (see Refs.~\cite{Tong2004, Singer2004, Pohl2011, Comparat2010} and references therein), thermal vapor microcells~\cite{Kuebler2010}, one-dimensional optical lattices~\cite{Viteau2011}, and on an atom chip~\cite{Tauschinsky2010}. In the field of quantum information processing, Rydberg atoms have been proposed as a medium for quantum computation~\cite{Jaksch2000, Saffman2010, Lukin2001, Mueller2009} and quantum simulations~\cite{Buluta2009, Weimer2010, Leung2011} because of their large dipole moments, their tunable long-range interactions and their strong coupling to microwave fields. In cavity quantum electrodynamics, Rydberg atoms passing through high-quality microwave cavities are at the heart of fundamental quantum optics and quantum information processing experiments~\cite{Walther2006,Haroche2006,Gleyzes2007}.

The long coherence times of Rydberg states make them attractive for hybrid systems in combination with solid-state devices operating in the quantum regime, such as circuit QED setups~\cite{Wallraff2004, Chiorescu2004}. Several approaches have been proposed to couple Rydberg atoms to superconducting circuits~\cite{Sorensen2004,Petrosyan2008,Petrosyan2009}. Also the interactions between Rydberg atoms and surfaces have been studied experimentally~\cite{Nordlander1996,So2011,Pu2013}. In experiments involving Rydberg atoms close to surfaces, the adsorption of atoms at the surface leads to inhomogeneous stray fields, which impose a severe limitation to the coherent manipulation of Rydberg atoms~\cite{Hattermann2012}, as has been described theoretically~\cite{Carter2011} and observed in experiments with alkali-metal atoms at room temperature~\cite{Obrecht2007, Tauschinsky2010, Carter2012,Carter2013a}. These harmful effects may be mitigated using microwave frequency dressing~\cite{Jones2013}. Such experiments are even more challenging if the surface is held at cryogenic temperatures because adsorption of atoms is enhanced \cite{Chan2013a} and typical surface cleaning procedures are not easy to implement.

In the experiments presented in this article, we use a supersonic beam of singlet He Rydberg atoms and let the Rydberg atoms pass within $500~\mu\text{m}$ of a thin-film normal metal or superconducting surface at cryogenic temperatures. Supersonic beams of Rydberg atoms offer advantages for studies of their interactions with surfaces and for experiments aiming at manipulating atoms and molecules near the surface of chips. They can be deflected, decelerated and even trapped using printed circuits~\cite{Hogan2012a,Allmendinger2013,Lancuba2013} while still maintaining the flexibility of choosing the atomic species and a well-defined initial state. In our case, the use of metastable helium allows us to mitigate the influence of stray fields originating from atoms from the atomic beam adsorbed onto the surface. Furthermore, we operate a specifically designed cryogenic experimental setup that minimizes adsorption of residual gases in the vacuum chamber onto the chip surface. Both measures allow for coherent manipulation of the Rydberg atoms close to the chip surface with pulses of microwave radiation. From the analysis of the observed patterns of Rabi oscillations, we were able to determine the distribution of stray electric and microwave fields above the sample surfaces.

\section{Experimental setup and measurement procedure} 
\label{sec:Experimental setup and measurement procedure}

\subsection{Experimental setup}
\label{sec:setup}

\begin{figure}[bth]
	  \centering
	  \includegraphics[width=86mm]{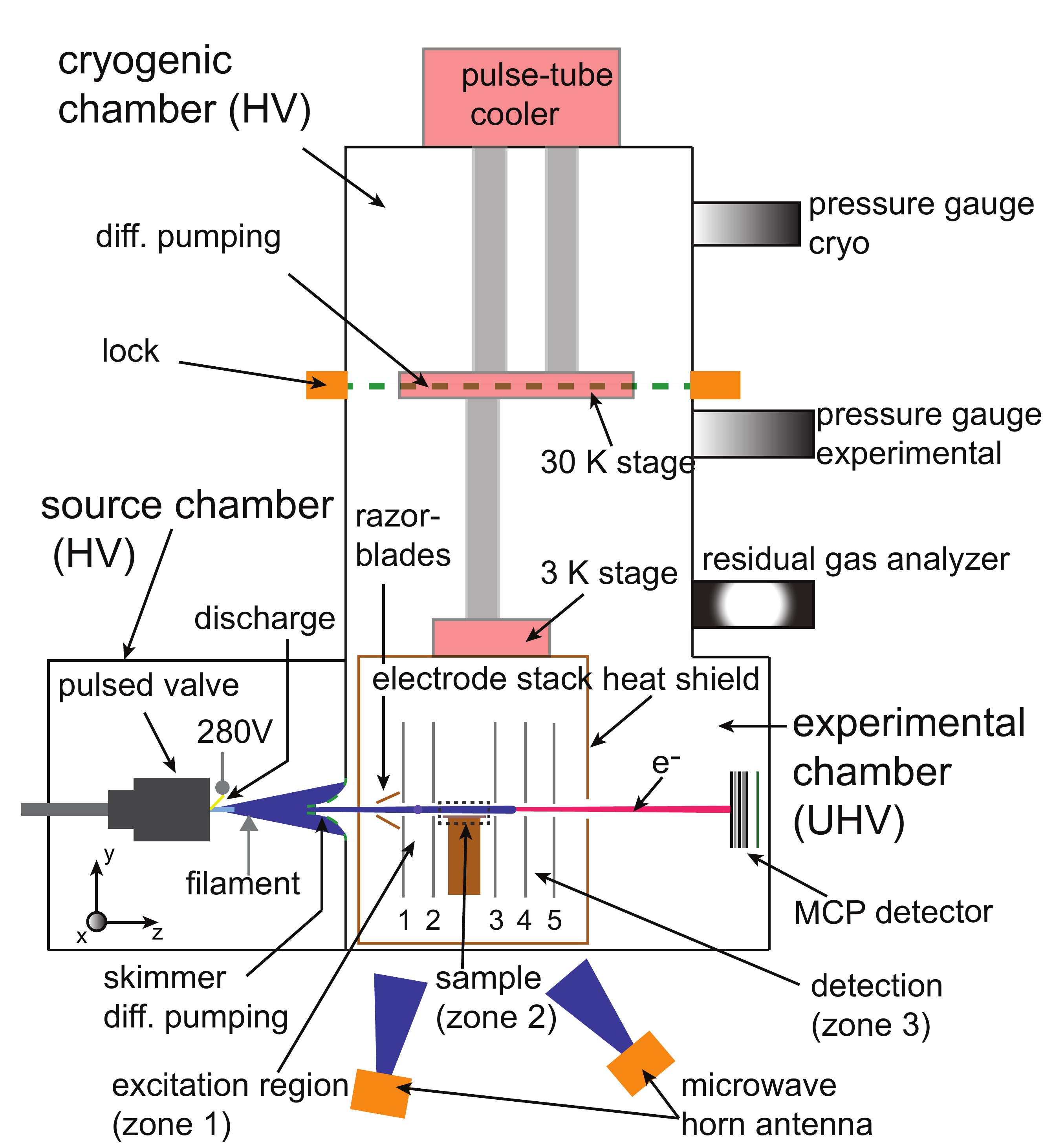}
	  \caption{Schematic view of the experimental setup consisting of the source chamber, the cryogenic chamber and the ultra-high-vacuum-compatible experimental chamber (see text for details).}
	  \label{fig:CoarseSetup}
	\end{figure}

The experimental setup consists of three differentially pumped vacuum chambers (see Fig.~\ref{fig:CoarseSetup}): The source chamber in which the atomic beam is generated; the cryogenic chamber holding the pulse-tube cooler; and the ultra-high-vacuum-compatible experimental chamber.
The pulse-tube cooler (Cryomech PT415RM) operates in a two-stage configuration. The first stage is held at $\approx30~\text{K}$ and separates the cryogenic chamber from the experimental chamber, the inner part of which is held at $3~\text{K}$ by the second cooling stage. Critical aspects of the cooling procedure are illustrated in Section~\ref{sec:Adsorption-free cooldown} and the technical details are provided in Appendix~\ref{sec:cooldownprocedure}.

In the source chamber, the supersonic beam of metastable $(1\text{s})^1(2\text{s})^1\, {}^1\text{S}_0$ He atoms, called He$^*$ hereafter, is generated by combining a pulsed supersonic expansion (repetition rate: $25~\text{Hz}$, pulse duration: $190 ~\mu\text{s}$), with a pulsed electric discharge seeded by electrons emitted from a hot tungsten filament, as described in Refs.~\cite{Halfmann2000, Hogan2012}. Each gas pulse contains about $10^{18}$ ground-state helium atoms, as estimated from pressure measurements and pumping rates, and about $10^{10}$ He$^*$ atoms. The steady-state pressure in the source chamber rises from $2\times10^{-7}~\text{mbar}$ to about $10^{-5}~\text{mbar}$ on average upon operation of the pulsed valve. To ignite the discharge, a $280$ V electric-tension pulse is applied for $20~\mu\text{s}$ in the $2~\text{mm}$ target region located immediately behind the nozzle orifice. 
	
The source-chamber and the experimental-chamber volumes are connected by the $1$-mm-diameter hole of a skimmer which allows for efficient differential pumping and selects the transversally coldest part ($0.01\%$ of all atoms) of the supersonic beam. The background pressure in the experimental chamber rises from $10^{-8}~\text{mbar}$ at room temperature and less than $5\times10^{-9}~\text{mbar}$ at $3~\text{K}$ (i.e. below the lower limit of the pressure gauge) to $5\times10^{-8}~\text{mbar}$ on average when the valve is pulsed. After the skimmer, the mean velocity of the He$^*$ beam is $1770~\text{m}/\text{s}$. The supersonic beam is further collimated in the $y$ direction (Fig.~\ref{fig:CoarseSetup}) by two parallel razor blades which are adjusted to the desired He$^*$ beam width, typically $250~\mu\text{m}$, and position in the experimental region (see Appendix~\ref{sec:distancecalibration} for details of the He$^*$ beam diagnostics).

The experimental region consists of three distinct zones. In the first, roughly $1000$ He$^*$ atoms are photoexcited to Rydberg states with a tunable, pulsed uv laser between two parallel disc-shaped metallic electrodes separated by $10~\text{mm}$ ($1$ and $2$ in Fig.~\ref{fig:CoarseSetup}). These plates have an outer diameter of $80~\text{mm}$ and a diameter of the inner hole of $5~\text{mm}$. The electrodes can be used to generate a homogeneous electric field at the spot where the atomic beam and the uv laser beam cross at right angles or to compensate stray fields in the zone located between electrodes $2$ and $3$.

In this second zone, the Rydberg beam propagates at a well-defined distance from the surface of the sample. Two types of samples have been used: a $500$-$\mu\text{m}$-thick $35~\text{mm}\times13~\text{mm}$ ($x\times z$) sapphire substrate coated with a $200$-nm-thick gold film, and a $30~\text{mm}\times12~\text{mm}$ sapphire substrate coated with a $150$-nm-thick superconducting NbTiN layer.

The third zone consists of three disc-shaped electrodes, separated by $10~\text{mm}$ (labeled $3,4$ and $5$ in Fig.~\ref{fig:CoarseSetup}), which are identical to those located in the first zone. Electrode $3$ is separated from electrode $2$ by $15~\text{mm}$. The Rydberg atoms are field ionized in this zone by applying a pulsed potential difference of $1.2~\text{kV}$ between either electrodes $3$ and $4$, or $4$ and $5$. The resulting electrons are accelerated toward a microchannel plate (MCP) detector in a chevron configuration connected to a phosphor screen for imaging. The electron signal is measured by capacitive coupling to the output (back) plate of the chevron stack.

Pulsed microwave radiation emitted from a horn antenna is used to induce Rydberg-Rydberg transitions in the three different zones by choosing an appropriate delay between the laser excitation pulse and the microwave pulse. The microwave radiation penetrates into the setup through the $5$-mm-diameter apertures also used for laser irradiation through the heat shields.

\subsection{Experimental procedure}
\label{sec:Life-time assisted spectroscopy}	

Rydberg states are produced by laser excitation of the $(1\text{s})^1(2\text{s})^1\, {}^1\text{S}_0$ metastable state of helium. Optical selection rules restrict transitions to the singlet $n\text{p}$ Rydberg series that converges to the $(1\text{s})^1{~}^2\text{S}_{1/2}$ ground state of $\text{He}^+$. We record laser excitation spectra of this Rydberg series by monitoring the electron signal produced by delayed pulsed field ionization as a function of the frequency of the uv laser. We detect the Rydberg states by field ionization in the third zone after a time of flight of $14~\mu\text{s}$, corresponding to a beam velocity of $1770~\text{m}/\text{s}$ and a distance of $\approx 25~\text{mm}$ between the laser excitation spot between electrodes $1$ and $2$ and the detection spot between electrode $3$ and $4$. The lifetime $\tau_n^{\text{p}}$ of the $n\text{p}$ series is estimated from $\tau_n^{\text{p}}=\tau_1^{\text{p}} {n^{*}}^3=\tau_1^{\text{p}} (n-\delta_p)^3$~\cite{Gallagher1994} with $\tau_1^\text{p}\cong3.8\times 10^{-11}~\text{s}$ determined from the measured lifetime of $\approx1.4~\mu\text{s}$ of the $34$p state and the quantum defect $\delta_p=0.98787$ of the p series~\cite{Drake1999}. This lifetime scaling law, when combined with the $(n^*)^{-3}$ dependence of the photoexcitation cross section, implies that the detection of Rydberg states below $n\approx40$ ($\tau_{40}^\text{p}\approx2.5~\mu\text{s}$) in the third zone is inefficient in the absence of stray electric fields [Fig~\ref{fig:stabilization}(b) in Section~\ref{sec:Adsorption-free cooldown}]. However, Rydberg states with $n<40$ are best suited for coherent manipulation with microwaves to study interactions with the sample~\cite{Hogan2012}; these states are hardly perturbed by stray electric fields, the $n\text{p}\leftrightarrow n\text{s}$ transition frequencies lie in the convenient range between $15-42~\text{GHz}$, and the p and s states can be field-ionized state selectively.

\begin{figure} [bth]
	  \centering
	  \includegraphics[width=78mm]{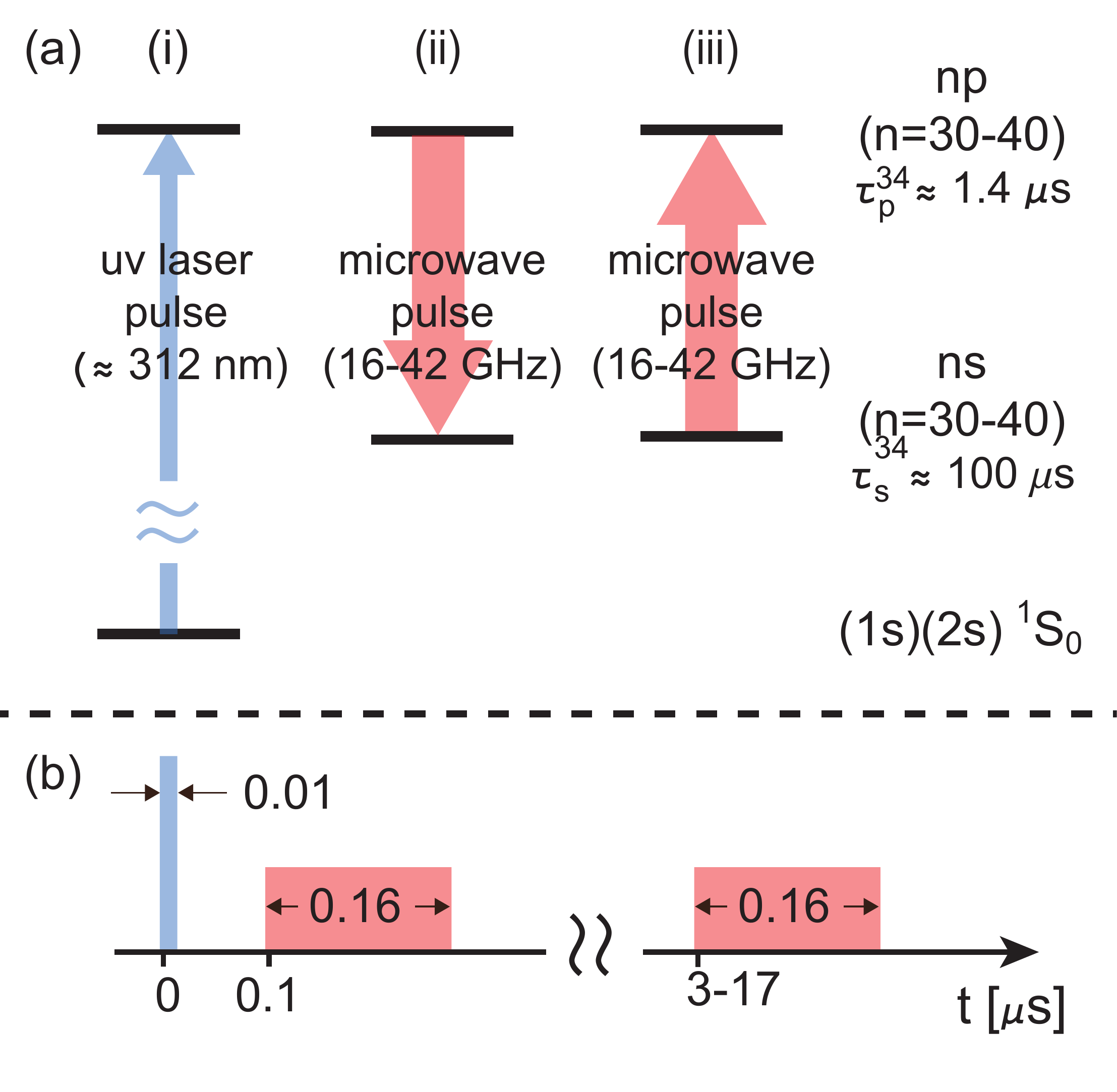}
	  \caption{(a) Schematic of the excitation sequence. The blue arrow indicates the laser-induced transition from the $(1\text{s})^1(2\text{s})^1\, {}^1\text{S}_0$ metastable state of He to a specific state of the $n\text{p}$ Rydberg series. The thick red arrows indicate transitions between $n\text{p}$ and $n\text{s}$ states stimulated by pulsed microwave radiation. (b) Time sequence and length of the radiation pulses.}
	  \label{fig:LevelSchemeAndMeasurementProcedure}
	\end{figure}

To overcome the detection inefficiency resulting from the short lifetimes of the $n\text{p}$ Rydberg states, we coherently transferred the Rydberg-state population from the $n\text{p}$ states to the nearest $n\text{s}$ states by applying a $160$-$\text{ns}$-long microwave pulse immediately after laser excitation [(ii) in Fig.~\ref{fig:LevelSchemeAndMeasurementProcedure}(a)]. The $n\text{s}$ states have lifetimes in the range of $65~\mu\text{s}$ to $165~\mu\text{s}$ for $n$ between $30$ and $40$, which is long enough to retain more than $80\%$ of the atoms in their excited state during their time of flight from the first to the third zone.

	In our experiments, the population in the $n\text{s}$ state can be transferred back to the short-lived $n\text{p}$ state by means of a second microwave pulse [(iii) in Fig.~\ref{fig:LevelSchemeAndMeasurementProcedure}(a)] at a variable time delay $t$ ranging from $3~\mu\text{s}$ to $17~\mu\text{s}$ with respect to the uv laser pulse. The decay of the $n\text{p}$ states during the remaining flight time to the third zone leads to a depletion of the pulsed field ionization signal.
	
\begin{figure} [!tbh]
\centering
\includegraphics[width=86mm]{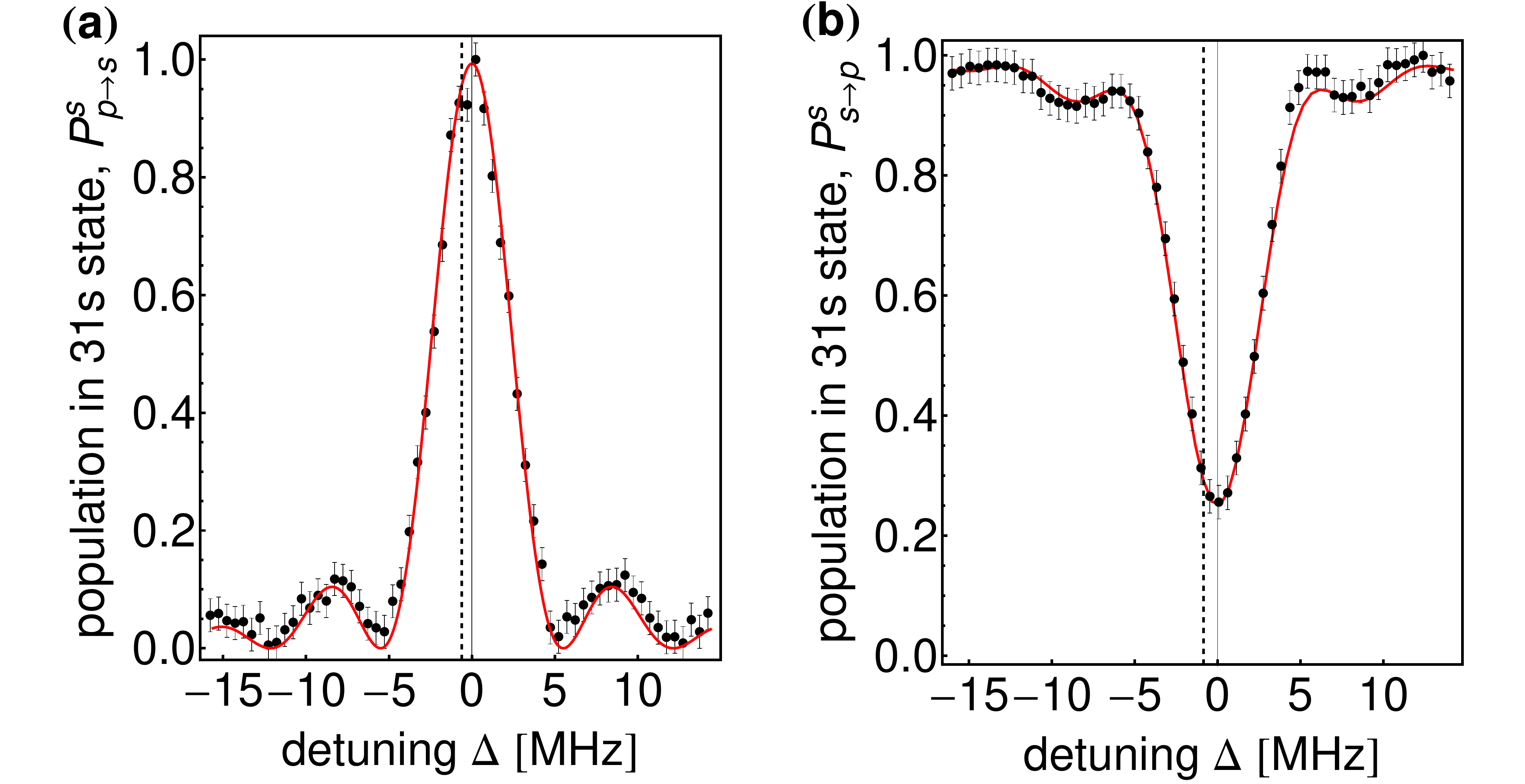}
\caption{(a) Spectrum of the $31\text{p}\rightarrow 31\text{s}$ transition of singlet helium recorded with a $160$-ns-long microwave pulse. The red line is a fit of $P_{\text{p}\rightarrow \text{s}}^\text{s}$, see Eq.~(\ref{eq:Rabifunction}). (b) Spectrum of the $31\text{s}\rightarrow 31\text{p}$ transition recorded with a second $160$-ns-long microwave pulse at a delay of $t=8~\mu\text{s}$. The red line is a fit of the expected s-state population $P_{\text{s}\rightarrow \text{p}}^\text{s}$, see  Eq.~(\ref{eq:Rabifunction2}), taking into account a Gaussian distribution of transition frequencies arising from the Stark effect caused by the stray field. The dashed vertical lines correspond to the calculated field-free frequency.}
\label{fig:freespaceRabi}
\end{figure}

We have recorded several spectra at room temperature with a single and with two microwave pulses, respectively (see Fig.~\ref{fig:freespaceRabi} for $n=31$), by detecting the pulsed field ionization signal in zone $3$ as a function of the microwave frequency. For these measurements no sample and no sample holder were inserted in region $2$. The data shown in Fig.~\ref{fig:freespaceRabi} is obtained by normalizing the signal to the maximum value and by subtracting a constant background signal which never exceeds $5\%$ of the total signal. The signal after the first pulse [Fig.~\ref{fig:freespaceRabi}(a)] is therefore approximately equal to the population of the s state, $P_{\text{p}\rightarrow\text{s}}^\text{s}$, at the time of field ionization ($18~\mu\text{s}$). This population, assuming that all atoms are initially in the p state, is described by the excitation probability
\begin{equation}
\label{eq:Rabifunction}
p(\Delta,\Omega,\Delta t)=\frac{\Omega^2}{\Delta^2+ \Omega^2} \sin{\left[\frac{\Delta t}{2} \sqrt{\Delta^2+\Omega^2}\right]}^2 \approx P_{\text{p}\rightarrow\text{s}}^\text{s},
\end{equation}
with the detuning $\Delta=\nu-\nu_0$ of the microwave frequency pulse, and the Rabi frequency $\Omega$. For a microwave pulse of duration $\Delta t=160~\text{ns}$ and $n=31$, a fit to Eq.~(\ref{eq:Rabifunction}) results in the Rabi frequency $\Omega/2\pi=2.958(84)~\text{MHz}$ and a center frequency $\nu_0=37245.775(28)~\text{MHz}$. The value $\nu_0$ is shifted by $614~\text{kHz}$ from the field-free transition frequency between the $31\text{s}$ and $31\text{p}$ states of $37245.161~\text{MHz}$ that is obtained by diagonalization of the single-particle Hamiltonian using the energy-dependent quantum defects of the helium atom~\cite{Zimmerman1979,Drake1999}. This shift results from the Stark effect induced by the stray electric field present in zone $1$, as discussed below.

Applying a second microwave pulse with variable frequency $\nu$ $8~\mu\text{s}$ after the first pulse of frequency $\nu_0$ transfers parts of the population back to the p state. The remaining population, $P_{\text{s}\rightarrow \text{p}}^\text{s}$, in the s state is modeled by the excitation probability $p$, see Eq.~(\ref{eq:Rabifunction}), averaged over a Gaussian distribution of transition frequencies $\rho(\Delta)=(\sqrt{2\pi}\sigma_\text{stray})^{-1}\exp{\left[-\Delta^2/(2\sigma_\text{stray}^{2}) \right]}$ caused by the slightly inhomogeneous stray electric field present at the time of the second microwave pulse,
\begin{equation}
\label{eq:Rabifunction2}
P_{\text{s}\rightarrow \text{p}}^\text{s}(\Delta')=1-\xi\int \rho(\Delta'-\delta) p(\delta,\Omega,\Delta t) \mathrm{d} \delta.
\end{equation}

For a $\Delta t=160~\text{ns}$ long pulse we obtain the Rabi frequency $\Omega/2\pi=3.18(23)~\text{MHz}$ and a center frequency $\nu'_0=\nu-\Delta'=37246.000(54)~\text{MHz}$ for the data shown in Fig.~\ref{fig:freespaceRabi}(b). The center frequency $\nu'_0$ is shifted by $840~\text{kHz}$ from the field-free frequency, indicating that the stray field is larger at this position than at the positions where the atoms are interacting with the first microwave pulse. We also find an incomplete population transfer probability of the second pulse, $\xi=90\%$, which can be explained by a non-uniform microwave excitation field across the atomic ensemble. The signal is not zero on resonance, because of the inhomogeneous broadening taken into account by the fit parameter $\sigma_\text{stray}=1.4(1)~\text{MHz}$.

\subsection{Electric field measurements}
\label{sec:ElectricField}

The resonance frequencies $\nu_0$ and $\nu'_0$ observed in Fig.~\ref{fig:freespaceRabi} provide information on the magnitude of the stray electric fields. At an electric field strength $F$ of less than $1~\text{V}/\text{cm}$, the Stark shift
\begin{equation} \label{eq:StarkFunction}
\Delta\nu_{\text{Stark}}=\frac{1}{2}\Delta\alpha F^2,
\end{equation}

of the $n\text{p}\leftrightarrow n\text{s}$ transition frequency is purely quadratic for $n<40$. For the $34\text{p}~\leftrightarrow~34\text{s}$ transition, the diagonalization of the Hamiltonian including the Stark shift~\cite{Zimmerman1979} leads to a value of $1078.03~\text{MHz}(\text{V}/\text{cm})^{-2}$ for the polarizability $\Delta\alpha$. To measure and compensate the stray electric fields, the method described in Ref.~\cite{Osterwalder1999} was used: the line shifts $\Delta\nu_{\text{Stark}}$ of the $34\text{p}\leftrightarrow 34\text{s}$ transition were measured for several intentionally applied potential differences between electrodes $1$ and $2$ (zone $1$), $2$ and $3$ (zone $2$), or $3$ and $4$ (zone $3$). In zone $1$, the field was estimated from spectra recorded with one microwave pulse, as illustrated in Fig.~\ref{fig:freespaceRabi}(a), whereas in zones $2$ and $3$, spectra obtained after applying two microwave pulses similar to the one in Fig.~\ref{fig:freespaceRabi}(b) were used.

\begin{figure}[tbh]
\centering
\includegraphics[width=86mm]{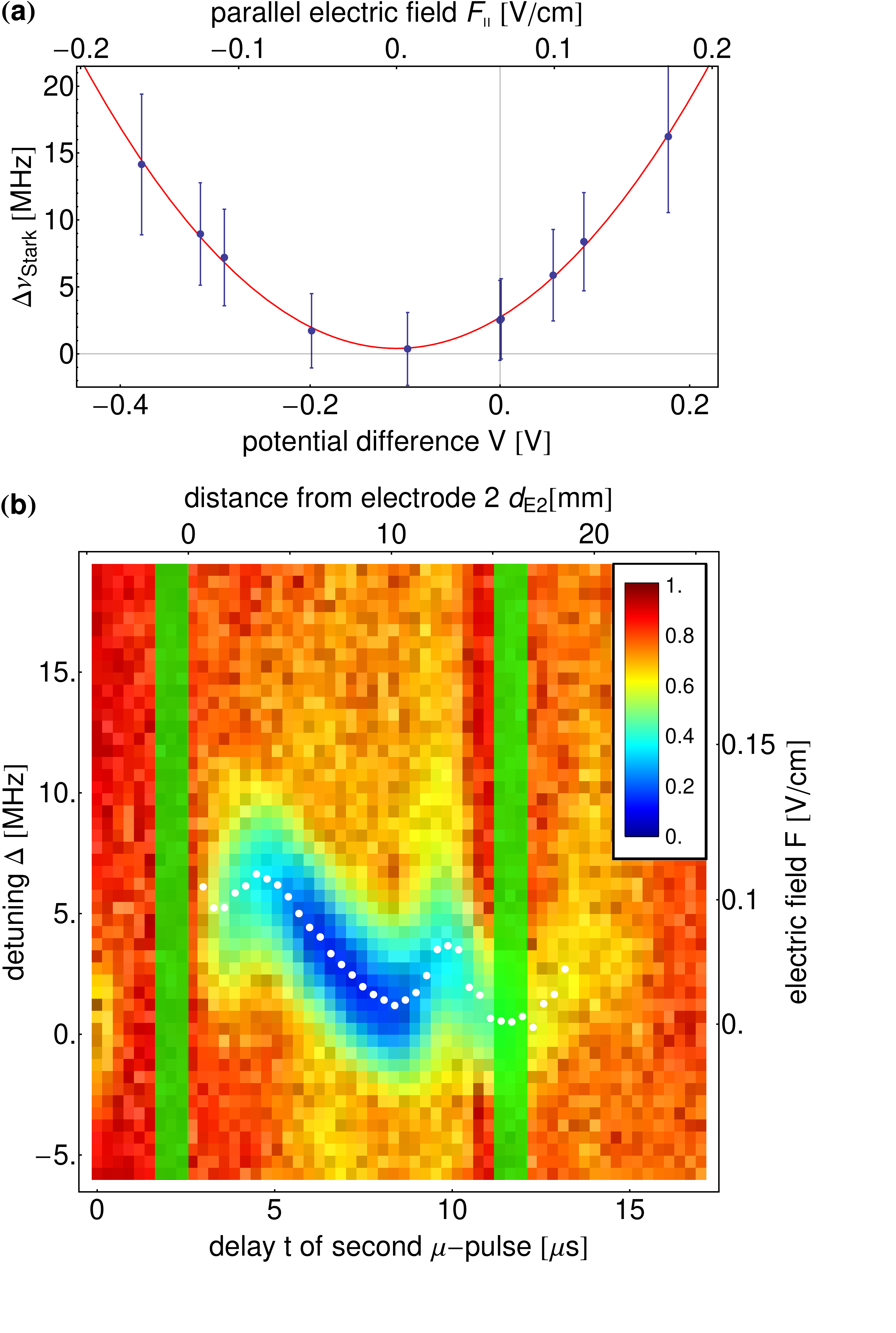}
	  \caption{(a) Measured transition frequency as a function of applied electric tension $V$ to electrode $3$ (bottom horizontal axis) and the corresponding parallel electric field $F_{\parallel}$ (top horizontal axis). The uncertainties of the transition frequencies are of the order of the size of the dots. The vertical bars correspond to the widths (FWHM) of the lines in MHz and the red line is a quadratic fit of the Stark shift in Eq.~(\ref{eq:StarkFunction}). (b) Population in the $34\text{s}$ state as a function of the time delay $t$ (bottom axis) and the detuning $\Delta$ from the field-free transition frequency (left axis). The blue areas indicate low $34\text{s}$ state population and the green areas show the positions of the electrodes $2$ and $3$. This provides a map of the stray electric field strength along the propagation axis (top axis) in the sample region with the right axis indicating the electric field $F$, determined from the fitted transition frequencies (white dots).}
	  \label{fig:PolarizationPlotTotal}
	\end{figure}

A typical stray-field measurement carried out $10~\mu\text{s}$ after the laser excitation in zone $2$ (without sample and sample holder) is displayed in Fig.~\ref{fig:PolarizationPlotTotal}(a). In this figure, $\Delta\nu_{\text{Stark}}$ is plotted against the applied potential difference V (bottom horizontal scale). The solid line represents a fit to the measured central frequencies using Eq.~(\ref{eq:StarkFunction}). The electric field $\vec{F}$ was decomposed into its components perpendicular ($F_{\perp}$) and parallel ($F_{\parallel}$) to the propagation direction ($z$ direction) of the atomic beam, $F^2 = F_{\perp}^2 + F_{\parallel}^2= F_{\perp,\text{stray}}^2 + (F_{\parallel,\text{stray}} - c V)^2$. The factor $c$ relates the potential difference $V$ between electrodes $2$ and $3$ to its parallel electric field component at the position of the atoms. With the value of $\Delta\alpha$ given above, one obtains stray-field components $F_{\parallel,\text{stray}}= 65(1)~\text{mV}/\text{cm}$ and $F_{\perp,\text{stray}} = 27(5)~\text{mV}/\text{cm}$. The extracted value of $c$ ($0.601(6)~\text{cm}^{-1}$) corresponds closely to the value of $0.615~\text{cm}^{-1}$ obtained from a two-dimensional finite-element calculation of the electric-field strength in zone $2$ for our electrode configuration. The $2\%$ discrepancy may originate from the uncertainty of the position of the atom cloud.

The position of the apex of the parabola in Fig~\ref{fig:PolarizationPlotTotal}(a) represents the potential difference for which the component of the stray field parallel to the beam-propagation axis is compensated, i.e., $F_\parallel=0$. A remaining shift at this position originates from a perpendicular component of the stray electric field, as well as further systematic shifts such as pressure shifts and ac Stark shifts. Because both the pressure shifts and the ac Stark shifts are negligible under our experimental conditions, we consider the perpendicular component of the stray field to be the dominant contribution to the shift of $414(160)~\text{kHz}$ of the zero-field resonance frequency.

By carrying out similar measurements for different delays between laser excitation and microwave pulse, we determine the local stray-electric-field strength along the propagation axis if the beam velocity is known. Such a measurement, carried out between electrodes $2$ and $3$ in zone 2 (without sample and sample holder), is depicted in Fig.~\ref{fig:PolarizationPlotTotal}(b). In this region, the stray field does not exceed $110~\text{mV}/\text{cm}$.

\section{Results}
\subsection{Adsorption of residual gas, charge build up and adsorption-free sample cooling procedure}
\label{sec:Adsorption-free cooldown}

The experiments presented here aim at coherently manipulating Rydberg atoms with microwave radiation emanating from superconducting chips at temperatures below $4~\text{K}$. Initial experiments indicated slow drifts of Rydberg-Rydberg microwave transition frequencies, slowly increasing line widths, and rapid decoherence whenever the Rydberg-atom beam propagated within $1~\text{mm}$ of the sample surface. These effects impede experiments on a timescale of only a few hours after cooling down the sample. They are caused by the adsorption of residual gas at the cold surface and the accumulation of charges in the matrix of frozen material, resulting in slowly increasing stray electric fields in zone $2$ of the experimental chamber (see also Ref.~\cite{Hogan2012}). Similar effects caused by adsorbed dipolar molecules and ions have been reported previously~\cite{Obrecht2007,Tauschinsky2010,Abel2011,Hattermann2012,Carter2012}.

\begin{figure}
	  \centering
	  \includegraphics[width=86mm]{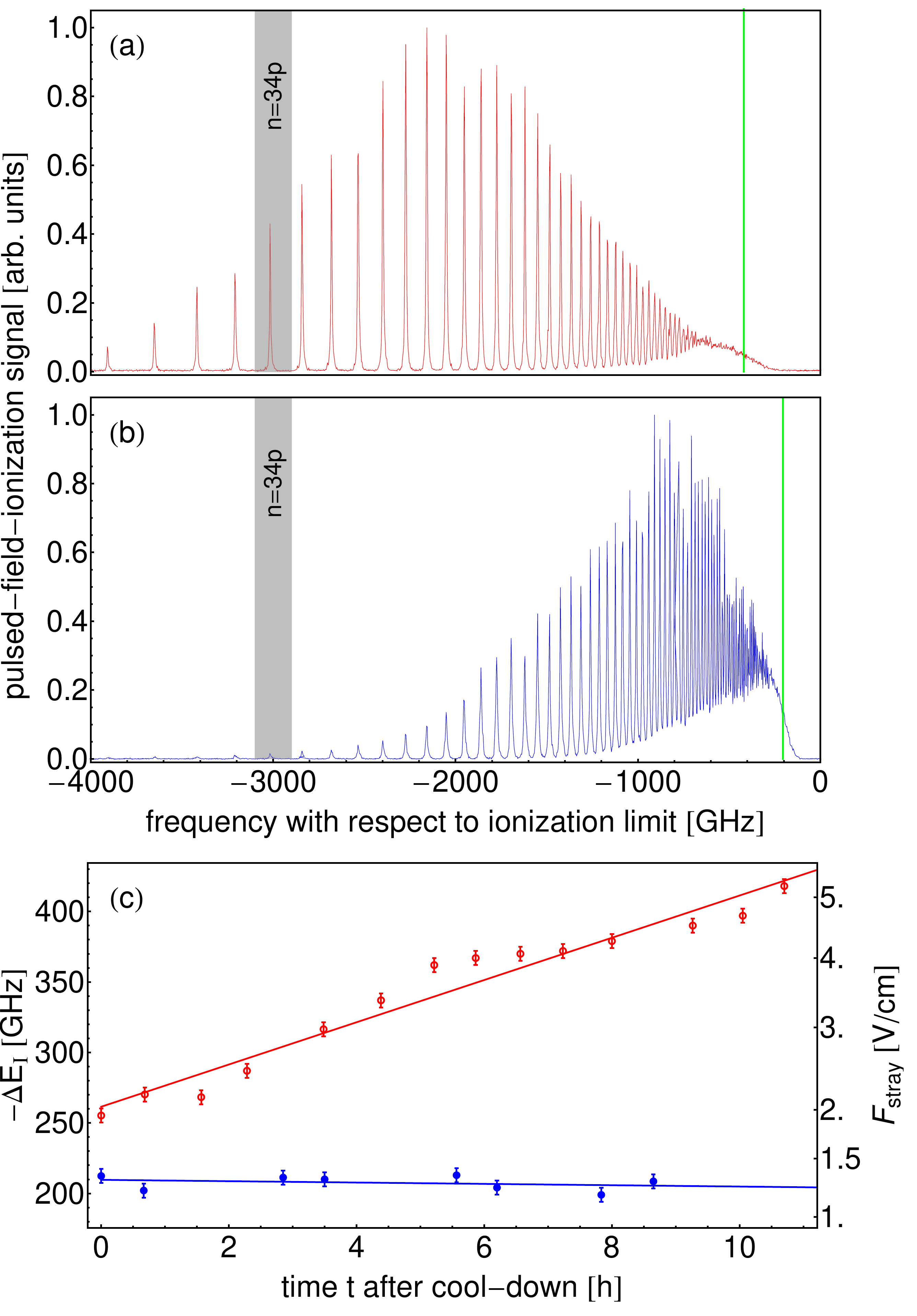}
	  \caption{Spectra of the $[\text{He}^+]n\text{p}\leftarrow(1s)^1(2s)^1{~}^1S_0$ Rydberg series of He recorded when (a) the residual gas adsorbed and charges build up in the matrix of frozen material, and when (b) the adsorption process onto the sample surface was suppressed. (c) Evolution of the field-induced shift of the ionization threshold ($-\Delta E_\text{I}$) versus time, with (full, blue circles) and without (open, red circles) suppression of adsorption of residual gas onto the sample surface.} 
	  \label{fig:stabilization}
\end{figure}

We have used the spectrum of the $[\text{He}^+]n\text{p}\leftarrow(1s)^1(2s)^1~{}^1\text{S}_0$ Rydberg series in the vicinity of the ionization threshold to monitor the build up of stray fields near the sample surface. After $10.5$ hours of experiments under conditions where charges could accumulate in the matrix of frozen material at the gold sample surface, the spectrum of this series [Fig.~\ref{fig:stabilization}(a)] markedly differs from the spectrum obtained under conditions where the adsorption of residual gas onto the sample surface was suppressed [Fig.~\ref{fig:stabilization}(b)] by means of the temperature-feedback procedure described in Appendix~\ref{sec:cooldownprocedure}. In both cases, the beam of Rydberg atoms excited in zone $1$ passed within $1~\text{mm}$ of the cold sample surface, and the spectra were recorded as explained in Section~\ref{sec:Life-time assisted spectroscopy}.

In Fig.~\ref{fig:stabilization}(a), the transitions to $n\text{p}$ Rydberg states below $n=50$ are strongly enhanced because the $l$-mixing~\cite{Chupka1993} of the initially prepared short-lived p Rydberg states induced by the stray electric fields increases the lifetimes. Moreover, the field-ionization signal rapidly decreases at frequencies less than $410~\text{GHz}$ (or $\approx12.3~\text{cm}^{-1}$) below the field-free ionization threshold ($0~\text{GHz}$), indicated by the vertical line in Fig.~\ref{fig:stabilization}(a). Using the well-known classical expression $\Delta E_\text{I}/(hc~\text{cm}^{-1})=-6.12\sqrt{F/(\text{V}/\text{cm})}$ \cite{Gallagher1994}, which relates the field-induced shift of the ionization energy $\Delta E_\text{I}$ to the field strength $F$, this observation indicates that the stray field along the beam-propagation axis reaches values up to $5~\text{V}/\text{cm}$ above the sample surface. In Fig.~\ref{fig:stabilization}(b), the weakness of the lines corresponding to transitions to $n\leq40$ Rydberg states and the observation of a field-ionization signal up to $200~\text{GHz}$ ($6~\text{cm}^{-1}$) below the field-free ionization threshold indicate that stray fields along the beam-propagation axis do not exceed values of $1.5~\text{V}/\text{cm}$.

By repeatedly recording spectra such as the one displayed in Fig.~\ref{fig:stabilization}(a) and determining from each spectrum the shift of the ionization threshold $\Delta E_\text{I}$, we have determined the gradual build-up of stray fields above the surface as the experiment progresses. $\Delta E_\text{I}$ is determined by fitting an error-function to the field-ionization signal close to the ionization limit and taking its inflection point. The results are depicted in Fig.~\ref{fig:stabilization}(c) as red open circles. Immediately after cooling the sample surface to $3~\text{K}$, the field-induced shift of the ionization threshold ($-\Delta E_\text{I}$) is $250~\text{GHz}$. It increases to $450~\text{GHz}$ at a rate of $15(1)~\text{GHz}/\text{hour}$, revealing that the maximal value of the stray field along the beam-propagation axis has increased to more than $5~\text{V}/\text{cm}$ after $11$ hours of continuous measurements.

To avoid the adsorption of residual gas on the sample surface while it is cooled down to $3~\text{K}$, a proportional-integral-derivative (PID) sample-temperature-control procedure was implemented, as described in Appendix~\ref{sec:cooldownprocedure}.
We monitor the partial pressures of the main components of the residual gas, i.e., H$_2$O, N$_2$, O$_2$ and CO$_2$, in the experimental chamber and maintain the temperature of the sample surface above the respective adsorption temperatures until the gases are fully adsorbed on the first cooling stage. In this way, the field-induced shift of the ionization threshold remains constant ($-0.5(6)~\text{GHz}/\text{hour}$) after cooling the sample surface to $3~\text{K}$, as shown by the full blue circles in Fig.~\ref{fig:stabilization}(c) and by the spectrum depicted in Fig.~\ref{fig:stabilization}(b), which was recorded after a continuous experimental run of $8.5$ hours.

\subsection{Measurement and compensation of stray electric fields above the sample surface}
\label{sec:compensation}
Stray fields emanating from the surface of the chip represent a major source of decoherence for Rydberg atoms close to the surface and need to be compensated. The measurements of the field-induced shifts of the ionization threshold presented in the previous section are well suited to detect and avoid adsorption onto the surface and the related build-up of stray fields. However, they do not provide enough information about the stray-field distribution above the surface along the beam-propagation axis for effective stray-field compensation.

To measure and minimize the stray fields above the sample surface after adsorption-free cooling to $3~\text{K}$, the method presented in Section~\ref{sec:ElectricField} was employed [Fig.~\ref{fig:PolarizationPlotTotal}(b)]. The method consists of determining the Stark shifts of microwave transitions between nearby Rydberg states, in the present case the $34\text{s}\leftrightarrow34\text{p}$ transition of He, using two microwave pulses separated by an adjustable delay. As explained in Section~\ref{sec:Life-time assisted spectroscopy}, the first pulse transfers the initial $34\text{p}$ population into the long-lived $34\text{s}$ state in zone $1$ of the experimental chamber. The second pulse is then used to determine the transition frequency, and therefore the Stark shift and the field strength, at positions which are uniquely defined by the velocity of the atomic beam and the time delay between the microwave pulses. The beam propagates at a mean distance of $250~\mu\text{m}$ above the gold-coated chip.

\begin{figure}[!t]
	  \centering
	  \includegraphics[width=89mm]{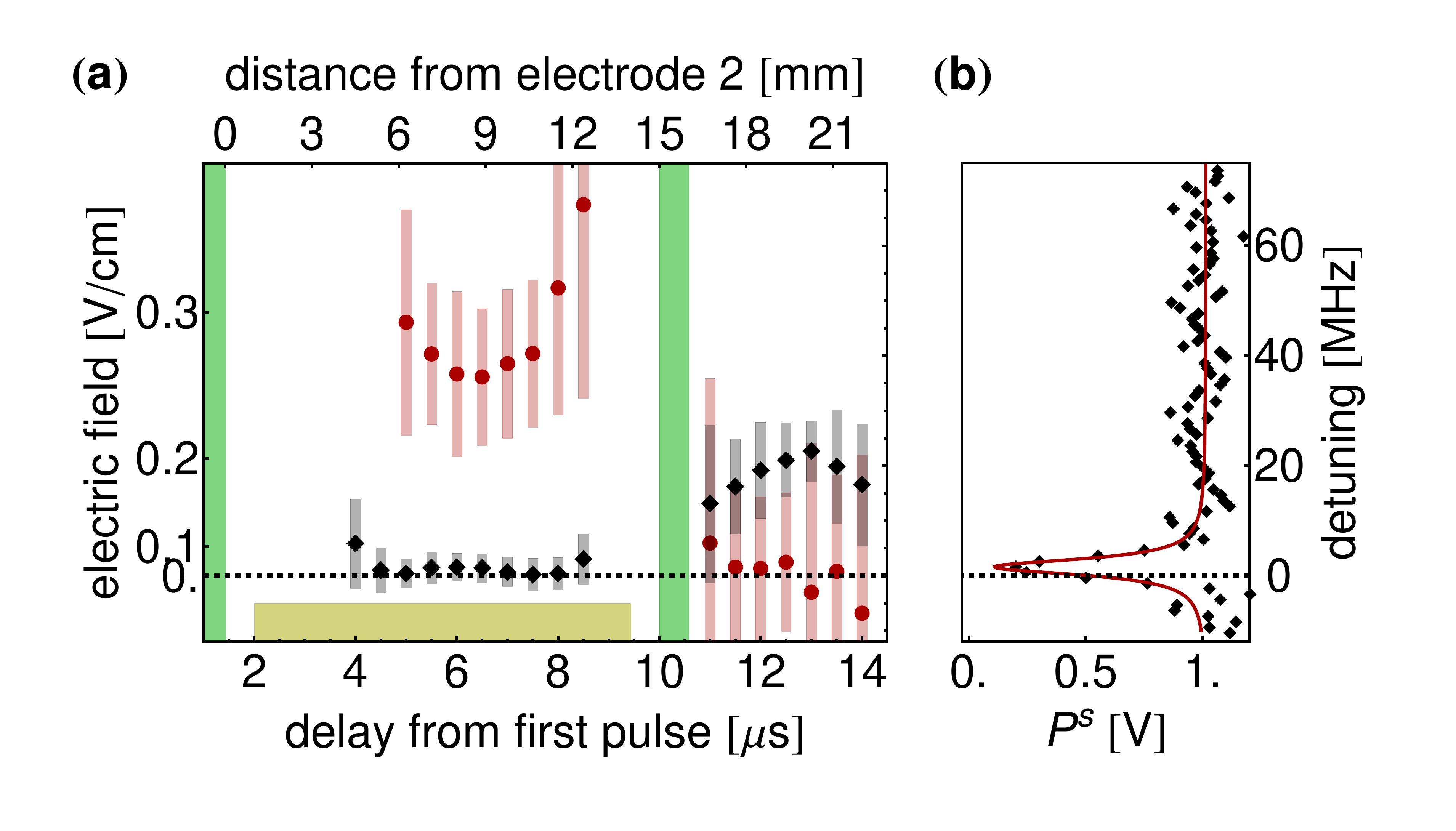}
	  \caption{(a) Measurement of stray fields above the chip surface (yellow horizontal box) using the $34\text{s}\leftrightarrow34\text{p}$ Stark shifts for atoms moving at a mean distance of $250~\mu\text{m}$ above the surface. The dots indicate the central positions of the transitions and the vertical bars indicate the corresponding width of the transitions. The data points indicate the measurement preformed without (red points) or with (black diamonds) stray-field compensation. Vertical green boxes indicate the position of electrodes $2$ and $3$. (b) Spectrum of the $34\text{s}\leftrightarrow34\text{p}$ transition with the stray-field compensation at a delay of $6~\mu\text{s}$.}
	  \label{fig:compensation}
	\end{figure}

The measured center positions of the spectral line and the extracted stray electric fields above the gold-coated chip  along the propagation axis of a beam are displayed in Fig.~\ref{fig:compensation}. The lower and upper horizontal axes in Fig.~\ref{fig:compensation}(a) represent the time delay between the two microwave pulses and the corresponding distance from electrode $2$, respectively. The es of the chip surface is indicated by a yellow box below the data points.

The data points in Fig.~\ref{fig:compensation}(a) correspond to the fields extracted from the observed center frequencies of the transitions, and the vertical bars indicate the range of fields accounting for their inhomogeneous widths. Prior to compensation (full circles), these widths range from $20~\text{MHz}$ ($\approx0.25~\text{V}/\text{cm}$) near the center of the chip ($6~\text{mm}$ away from electrode $2$), to $70~\text{MHz}$ ($\approx0.36~\text{V}/\text{cm}$) near the edge of the chip ($12.5~\text{mm}$ from electrode $2$). The data reveal a rapid increase of the stray field and field inhomogeneities near the edges of the chip such that no transitions are observed at delay times of less than $5~\mu\text{s}$ and between $8.5$ and $11~\mu\text{s}$. At times beyond $11~\mu\text{s}$, the Rydberg atoms are located in the field-free region beyond electrode $3$, in which the microwave power is above the saturation threshold of the transition such that it contributes to the observed linewidths.

To compensate the measured stray fields above the chip surface, we apply the potentials $V_2$ and $V_3$ (with respect to a common ground) to electrodes $2$ and $3$, respectively. This results in a dominant component of the electric-field vector pointing in the direction perpendicular to the surface near the center of the chip ($y$ direction) and stray-field inhomogeneities near the edges of the sample surface as calculated in a finite-element simulation. The position along the beam-propagation axis at which the electric field is exactly perpendicular to the surface can be adjusted by applying different potential differences between electrodes $2$ and $3$. 

Optimal stray-field compensation was achieved above the surface over a distance of $\approx6~\text{mm}$ along the beam-propagation axis by setting $V_2=300~\text{mV}$ and $V_3=255~\text{mV}$ [Fig.~\ref{fig:compensation}(a); black diamonds]. The observed Stark shifts are reduced to less than $1.5~\text{MHz}$ at distances between $5$ and $11~\text{mm}$ from electrode $2$, corresponding to residual fields of less than $55~\text{mV}/\text{cm}$. These are of the same order of magnitude as the residual fields ($F_\perp$) observed after stray-field compensation in the absence of the sample [Fig.~\ref{fig:PolarizationPlotTotal}(b)]. The linewidths of the microwave transitions are in the range of $5-6~\text{MHz}$, limited by the bandwidth of the pulsed microwave radiation [Fig.~\ref{fig:compensation}(b)]. Because of the compensation potential $V_3$ applied to electrode $3$, the field in zone $3$ of the experimental chamber is no longer zero.

By changing the potentials $V_2$ and $V_3$ to $200~\text{mV}$ and $255~\text{mV}$, respectively, the region in which the stray field is compensated can be increased to almost $9~\text{mm}$ and displaced by $\approx2~\text{mm}$ along the beam-propagation axis (not shown) at the cost of optimal compensation. In this case, the observed Stark shifts after stray-field compensation are smaller than $6~\text{MHz}$. However, it was not possible to compensate the stray fields above the entire chip surface because of the stray-field inhomogeneities near the edges of the surface. Since a similar behaviour is observed for the superconducting NbTiN surface, these results imply the possibility of coherent manipulation of Rydberg atoms near superconducting surfaces for several microseconds.

\subsection{Coherent Rydberg-Rydberg population transfer as probe of stray electric and microwave fields above chip surfaces}
\label{sec:coherentRabis}

The suppression of adsorption at the sample surface and the compensation of stray electric fields to levels much below $1~\text{V}/\text{cm}$ demonstrated in our experiments make it possible to observe coherent population transfer between the $34\text{s}$ and $34\text{p}$ Rydberg states of He induced by microwave radiation pulses at a distance of less than $500~\mu\text{m}$ from the sample surface. Rabi oscillations were measured in dependence on the detuning $\Delta$ from the field-free resonance frequency of $27965.77~\text{MHz}$ by either varying the microwave power at constant pulse length, or by varying the microwave pulse length at constant power. From the measured s-state population informations about the remaining electric field distribution close to the surface is extracted. The experiments presented in this section were carried out with both the gold-coated surface and the surface coated with a superconducting NbTiN layer, both held at 3~\text{K}. In both cases, the atomic beam was confined by the razor blades to a mean distance of $250~\mu\text{m}$ from the surface with an estimated full width at half maximum of $250~\mu\text{m}$.

In the first set of experiments carried out above the gold surface, we apply $160$-ns-long microwave pulses at a delay of $6~\mu\text{s}$ such that the excited atoms were located in zone $2$ at a distance of $\approx8~\text{mm}$ from electrode $2$ [Fig.~\ref{fig:compensation}(a)]. To compensate the stray field, we applied electric potentials of $V_2=200~\text{mV}$ and $V_3=255~\text{mV}$ to electrodes $2$ and $3$, respectively, resulting in a Stark shift of $\approx6~\text{MHz}$ corresponding to a residual field of $\approx110~\text{mV}/~\text{cm}$ at the position of the atoms. For detunings $\Delta$ varying between $-5~\text{MHz}$ and $15~\text{MHz}$ from the field-free resonance frequency, the population transfer was monitored by pulsed field ionization in zone $3$ for applied microwave powers $P_\mu$ ranging from $0$ to $5~\text{mW}$. The microwave power was set at the microwave source and corresponds to microwave field amplitudes $F_\mu$ ranging from $0$ to $0.4~\text{V}/\text{cm}$. $F_\mu=\hbar\Omega_\text{max}/\mu_\text{el}$ was determined using the known transition electric-dipole moment $\mu_{\text{el}}$ of the $34\text{s}\leftrightarrow34\text{p}$ transition and the Rabi frequency $\Omega_{\text{max}}$ observed at the maximal amplitude on resonance.

	\begin{figure} [tbh]
	  \centering
	  \includegraphics[width=86mm]{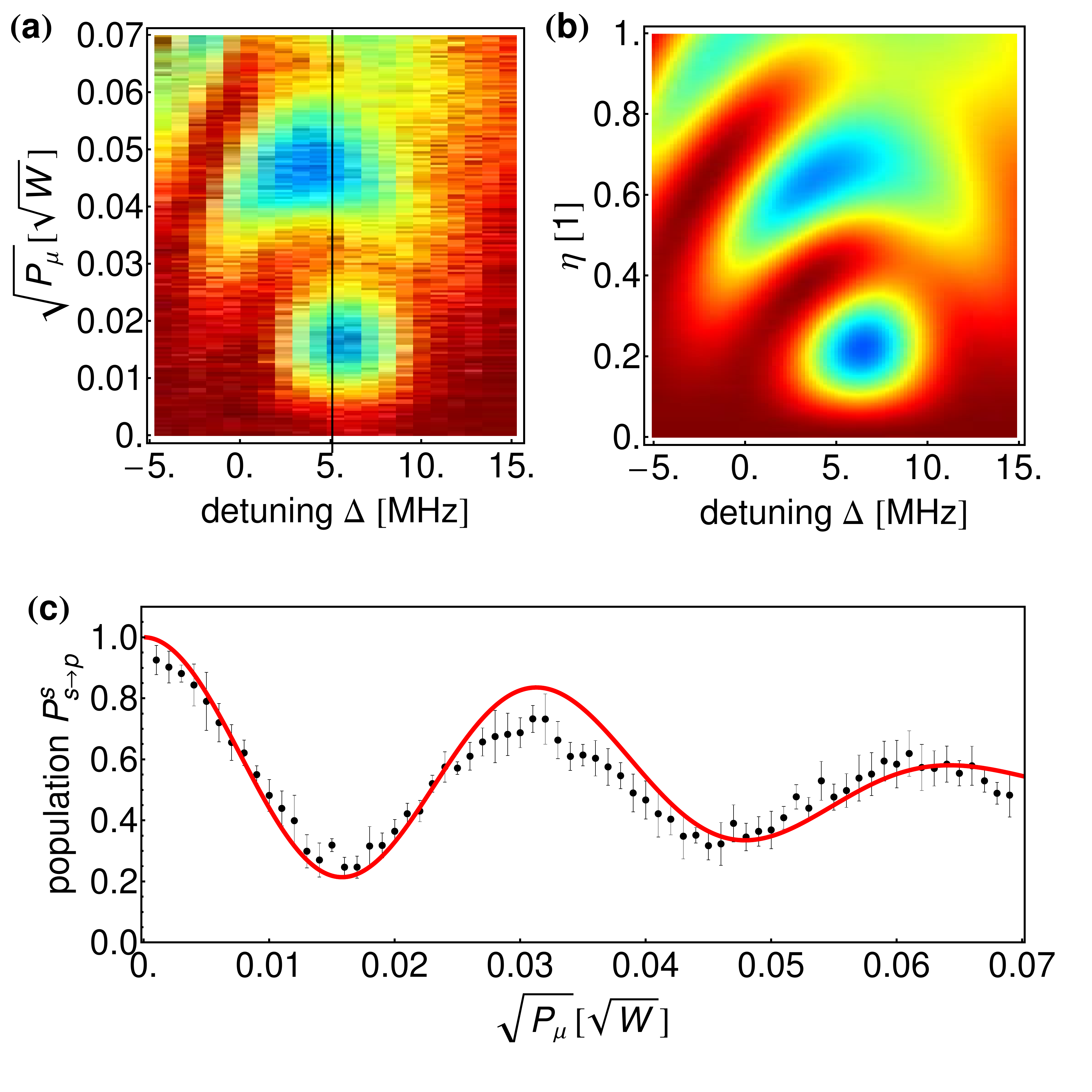}
	  \caption{(a) Rabi oscillations measured as a function of microwave power $P_\mu$ for different detunings $\Delta$ from the field-free resonance frequency of the $34\text{s}\leftrightarrow34\text{p}$ transition of singlet He, for a Rydberg atom cloud moving at a mean distance of $250~\mu\text{m}$ above the gold surface. (b) Corresponding simulation as described in the text. $\eta$ is varied linearly from $0$ to $1$ when the experimentally set amplitude at the microwave source is varied as the square root of the power between $0$ and $5~\text{mW}$. In (a) and (b), the color scale ranges from $0.14$ (blue) to $1$ (dark red). (c) Measured (black dots) and simulated (red line) Rabi oscillations at a detuning of $5~\text{MHz}$ indicated by the dashed vertical line in (a). The vertical scale is normalized to the maximum field-ionization signal of the $34\text{s}$ Rydberg state observed at a detuning $\Delta=-4.9~\text{MHz}$ and microwave powers below $2.5~\text{mW}$.  For better visibility the number of points was reduced by taking the average of $5$ consecutive datapoints.}
	  \label{fig:RabiSimulationTotal}
	\end{figure}
The experimental data in Fig.~\ref{fig:RabiSimulationTotal}(a) show two regions of parameters $\Delta$ and $P_\mu$ in which the population transfer to the short-lived $34\text{p}$ state is efficient. The first one is for detunings around $6.5~\text{MHz}$ and microwave powers around $0.3~\text{mW}$, and the second one is for detunings around $4~\text{MHz}$ and microwave powers around $2.5~\text{mW}$. The experimental data reveal a pronounced asymmetry which is caused by inhomogeneities of the stray electric and microwave fields across the volume of the $34\text{s}$ Rydberg-atom cloud. For the same reason the visibility of the Rabi oscillations is less than unity, as exemplified by the measurement at $\Delta=5~\text{MHz}$ [Fig.~\ref{fig:RabiSimulationTotal}(c)].

To simulate the experimental data, we refine Eq.~(\ref{eq:Rabifunction2}) by introducing a parametrization of the stray electric field and microwave field distribution over the atom cloud in order to derive the distribution $\nu(y)$ of the atomic transition frequencies from the spatial distribution of the atoms $\rho(y)$. The s-state population is then calculated using

\begin{equation}
\begin{split}
&P_{\text{s}\rightarrow\text{p}}^{\text{s}}(\Delta=\nu-\nu_0,\eta) =	\\
&1-\int_0^{500~\mu\text{m}}{ \rho(y) p\left(\nu-\nu(y),\Omega(\eta, y),\Delta t\right)dy,}
\end{split}
\end{equation} 
where we model the spatial distribution of the $34\text{s}$ Rydberg atoms in the $y$-direction perpendicular to the surface by a Gaussian distribution $\rho(y)$ with maximum at $y_\text{max}=250~\mu\text{m}$ and a full width at half maximum of $250~\mu\text{m}$. The integration limits are dictated by the positions of the razor blades. We assume $\nu$ and $\Omega$ to vary linearly over the atomic ensemble with the distance $y$ from the surface according to
  \begin{equation}
  \label{Fieldele}
     F_{\text{stray}}(y) = F^{0}_{\text{stray}} + F^{1}_{\text{stray}}(y-y_\text{max}), 
   \end{equation}
and
	\begin{equation}  
	\label{Fieldmic} 
     F_{\mu}(y) =\eta\left(F^{0}_{\mu} + F^{1}_{\mu}(y-y_\text{max})\right), 
	\end{equation}
respectively. $\eta$ corresponds to the microwave amplitude varying linearly from $0$ to $1$ when the experimentally set amplitude at the microwave source is varied as the square root of the power between $0$ and $5~\text{mW}$. Eqs.~(\ref{eq:StarkFunction}),(\ref{Fieldele} and (\ref{Fieldmic} imply a direct correspondence between the distance $y$ of an atom from the sample surface, its transition frequency $\nu(y)$ and the Rabi frequency $\Omega(\eta,y)\propto\mu_{\text{el}}(F_{\text{stray}}(y))F_{\mu}(\eta,y)\propto y^2$, because for the relevant fields $\mu_\text{el}\propto F_{\text{stray}}$, and $F_{\text{stray}}$ and $F_{\mu}$ depend linearly on $y$. Good agreement with the experimental data is obtained for $F^{0}_{\text{stray}}=0.110(5)~\text{V}/\text{cm}$, $F^{1}_{\text{stray}}=-1.2(5)~\text{V}/\text{cm}^2$, $F^{0}_{\mu}=0.158(9)~\text{V}/\text{cm}$, and $F^{1}_{\mu}=1.23(4)~\text{V}/\text{cm}^2$, see Eqs.~(\ref{Fieldele}) and (\ref{Fieldmic}). The dc electric field decreases with distance from the gold surface whereas the microwave field increases with distance from the gold surface.

The good agreement between the experimental and simulated plots in Fig.~\ref{fig:RabiSimulationTotal} indicates that the measurement of coherent population transfer provides information not only on the stray-field distribution, but also on the distribution of microwave fields above the sample surface.

In a second set of experiments, carried out above a superconducting NbTiN surface, the microwave pulse length $\Delta t$ was varied between $0$ and $500~\text{ns}$ at a constant source power ($\approx 31.6~\text{mW}$) for detunings varying between $-2$ and $8~\text{MHz}$. The microwave pulses were applied at a delay of $5.5~\mu\text{s}$. To compensate the stray field, we applied the potentials $V_2=110~\text{mV}$ and $V_3=200~\text{mV}$ to electrodes $2$ and $3$, respectively, resulting in a Stark shift of $\approx2~\text{MHz}$ at the position of the atoms. 
\begin{figure}[!htbp]
	  \centering
	  \includegraphics[width=86mm]{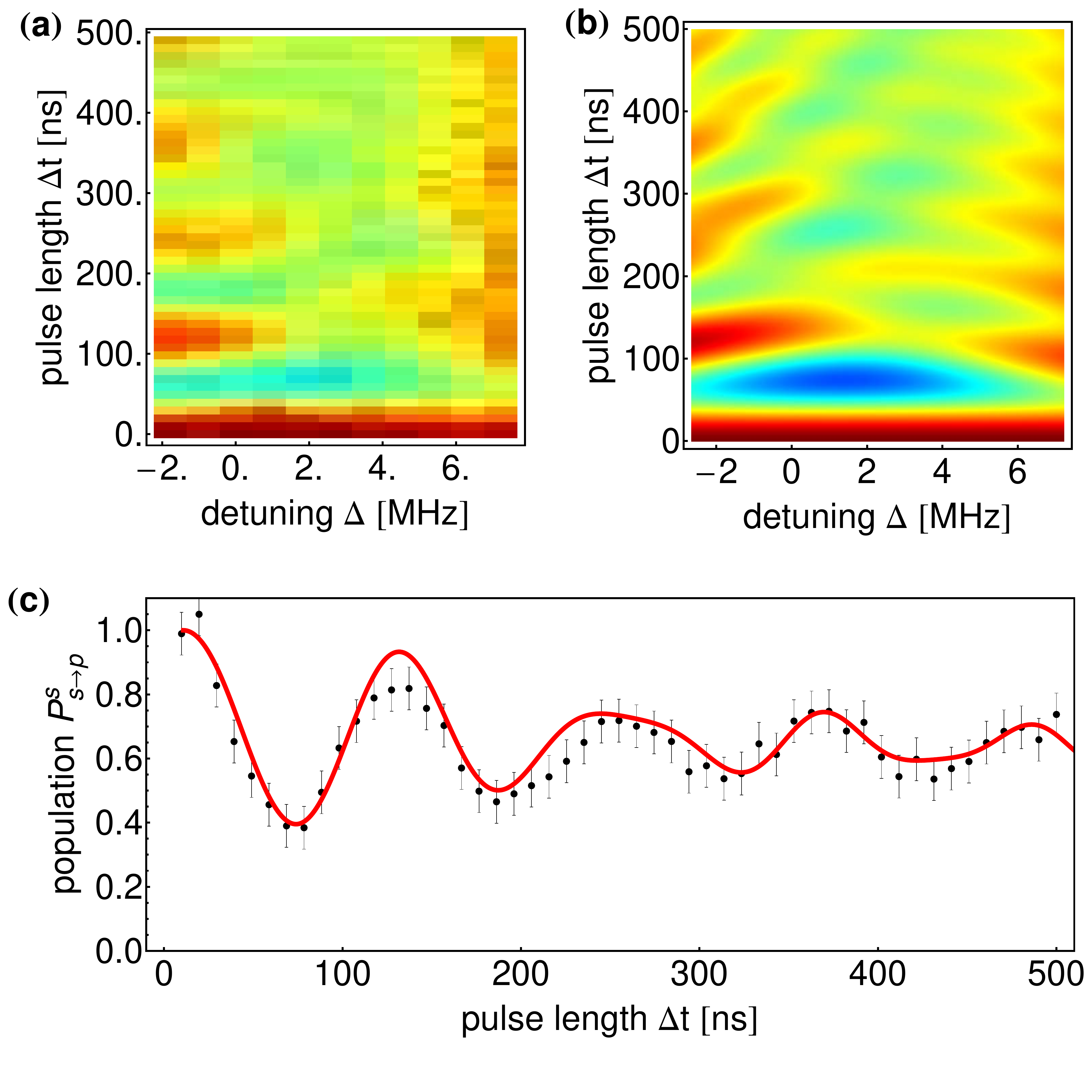}
	  \caption{(a) Population $P_{\text{s}\rightarrow\text{p}}^{\text{s}}$ as a function of microwave pulse length $\Delta t$ versus detuning $\Delta$ from the field-free resonance frequency of the $34\text{s}\leftrightarrow34\text{p}$ transition of singlet He. The Rydberg atom cloud is moving with a center of mass position of about $250~\mu\text{m}$ above the NbTiN surface. (b) Corresponding simulation as described in the text. (c) Observed (black dots) and simulated (red line) Rabi oscillations at a detuning of $-2~\text{MHz}$.}
	  	  \label{fig:RabiSimulationTotalTime}
	\end{figure}
As in the first set of experiments the experimental data shown in Fig.~\ref{fig:RabiSimulationTotalTime}(a) exhibit an asymmetry in frequency with respect to $\Delta=0$ and a reduced visibility of the Rabi oscillations, as illustrated for the measurement at $\Delta=-2~\text{MHz}$ in Fig.~\ref{fig:RabiSimulationTotalTime}(c).

At the velocity of $1770~\text{m}/\text{s}$ of the supersonic beam, the atoms move almost $1~\text{mm}$ during the longest pulses ($500~\text{ns}$) used in these experiments. Hence, it was necessary to also consider the gradient of the microwave field strength along the beam-propagation axis in the simulations. This was done by introducing a time-dependent averaged Rabi frequency
\begin{equation}
\label{eq:timedependentrabi}
\bar{\Omega}(y,\Delta t)\approx\frac{1}{\Delta t}\int_{0}^{\Delta t}\Omega(t)dt\approx\Omega(y)\left(1+m\Delta t\right).
\end{equation}

The simulated data in Fig.~\ref{fig:RabiSimulationTotalTime}(b) correspond to the s state population determined by 
\begin{equation}
\begin{split}
&P_{\text{s}\rightarrow\text{p}}^{\text{s}}(\Delta=\nu-\nu_0,\Delta t) =	\\
&1-\int_0^{500~\mu\text{m}}{ \rho(y) p\left(\nu-\nu(y), \bar{\Omega}(y,\Delta t),\Delta t\right)dy,}
\end{split}
\end{equation}
with $F^{0}_{\text{stray}}=0.040(15)~\text{V}/\text{cm}$, $F^{1}_{\text{stray}}=-3.5(5)~\text{V}/\text{cm}^2$, $F^{0}_{\mu}=0.099(3)~\text{V}/\text{cm}$, $F^{1}_{\mu}=2.0(2)~\text{V}/\text{cm}^2$ and $m=0.5/(500~\text{ns})$. The value of $m$ implies a microwave field gradient of $\approx 0.58(2)~\text{V}/\text{cm}^2$ along the beam-propagation axis. In Fig.~\ref{fig:RabiSimulationTotalTime}(c) the experimental data (dots) recorded at $\Delta=-2\text{MHz}$ and the corresponding simulation (red line) are compared. This data set illustrates the good agreement between experimental results and simulation.

In both measurements, the extrapolation of the microwave field to $y=0$ yields a value of $\approx0.12~\text{V}/\text{cm}$, independent of the chosen material, gold or superconducting NbTiN. In fact, the distribution of the microwave field in the sample region (zone $2$) is fully determined by the geometry of the setup. In our experiment the microwave radiation penetrates into zone $2$ of the experimental chamber through the $5$-mm-diameter holes of electrodes $2$ and $3$ and has non-zero amplitude in the chip, which is located only $1~\text{mm}$ below the axis of the holes. Whether the microwave field vanishes at the sample surface or not, depends on the field distribution in zone $2$ and is independent of the conductivity of the surface in a first approximation.

\section{Conclusions}

An atomic beam of singlet He Rydberg atoms has been manipulated coherently close to gold and superconducting surfaces at cryogenic temperatures ($3~\text{K}$). Collimation of the beam with a skimmer and a pair of razor blades in combination with beam imaging techniques enabled the formation of a narrow beam (full width at half maximum of about $250~\mu\text{m}$) with a mean distance that can be adjusted to approximately $250~\mu\text{m}$ above the cold surface.

Differential pumping between the vacuum chambers and a cooling procedure that holds the temperature of the sample surfaces at set temperatures, chosen to be higher than the adsorption temperatures of residual gases with measurable partial pressures, enabled the suppression of adsorption of residual gas and the build-up of stray electric fields near the surfaces.

To overcome sensitivity losses imposed by the short lifetimes of singlet $n\text{p}$ Rydberg states of He with $n$ between $30$ and $40$ prepared from the $(1\text{s})^1(2\text{s})^1\, {}^1\text{S}_0$ metastable level, a microwave pulse of well controlled length and power was used to coherently transfer the population to the neighboring long-lived $n\text{s}$ levels immediately after laser excitation. After an adjustable delay, the population was converted back to the original $n\text{p}$ level by a second  microwave pulse which is used to probe the Rydberg atoms as they propagated above the sample surface. With this two-pulse technique, the stray electric fields above the surfaces were mapped out along the beam-propagation axis. The fields emanating from the surface were compensated to a level below $100~\text{mV}/\text{cm}$ over a distance of $6~\text{mm}$ directly above the cold surface.

Rabi oscillations between the $34\text{p}$ and the $34\text{s}$ levels, for a He beam propagating $250~\mu\text{m}$ above a gold and a superconducting NbTiN surface, were recorded by changing the frequency and either the amplitude or the length of  microwave radiation pulses. The observed patterns of Rabi oscillations provide information on the distributions of microwave and stray electric fields above the surface.

The coherence times achieved for the Rydberg atom sample in the immediate vicinity of superconducting surfaces, and the ability to manipulate Rydberg atoms with microwave radiation pulses above these surfaces are prerequisite for experiments in which Rydberg states are coupled coherently to superconducting resonators embedded in solid-state devices.

\textbf{Acknowledgements}: We thank Dr. S. D. Hogan, (University College London) for his contributions to the initial phase of the project and for fruitful discussions as well as A. Hambitzer for determining the atom number in the source chamber and the velocity of the atomic beam. This work is supported by the NCCR QSIT of the Swiss National Science Foundation, and the ERC.

\appendix

\section{Avoiding adsorption of residual gas on surfaces in the experimental chamber}

\label{sec:cooldownprocedure}

In our earlier set of experiments~\cite{Hogan2012}, the adsorption of residual gas onto the sample surface led to the build up of charges in the matrix of adsorbed material which caused rapidly increasing inhomogeneous stray electric fields, line shifts to higher frequencies and line broadenings, as discussed in Section~\ref{sec:Adsorption-free cooldown}. To avoid these effects we had limited the operating temperature to $T>100~\text{K}$ in these experiments. To minimize the adsorption onto the sample surface and to reach workable temperatures below $4~\text{K}$ the apparatus and cooling procedure were modified in the present experiments so that the residual gas adsorbed predominantly at the first stage and not at the sample holder and sample surfaces. 

First, the cryogenic chamber was designed to avoid gas particle diffusion into the experimental chamber without colliding with a cold surface. Second, the sequence with which the different elements were cooled was optimized and the actual cooling process was monitored by measuring the temperature of the sample holder and the partial pressures of the relevant gaseous species (H$_2$O, CO$_2$, N$_2$ and O$_2$) with a residual gas analyzer (RGA). In fact, most of the residual gas enters the chamber through the viton o-rings of the cryogenic chamber, which reaches a base pressure of $5\times10^{-7}~\text{mbar}$ at room temperature and less than $5\times10^{-9}~\text{mbar}$ (i.e., below the lower limit of the pressure gauge) at $32~\text{K}$. The key to avoid adsorption on the sample holder is to keep its temperature above the adsorption temperature of the different gaseous components as long as these have measurable partial pressures (i.e., $>5\times10^{-10}~\text{mbar}$) in the experimental chamber. This approach leads to the gas adsorbing predominantly onto the first stage. 

Before starting the cooling procedure, pairs of reference partial-pressure and temperature values ($p_i^{\text{ref}}$, $T_i^{\text{ref}}$) are defined for all relevant gaseous species ($i=1-4$ for H$_2$O, CO$_2$, N$_2$ and O$_2$) as indicated in Table~\ref{tab:setpoints}. While continuously monitoring the partial pressures, the temperature of the sample is maintained with a cartridge heater at the maximal $T_i$ value for which the condition $p_i^{\text{measured}}>p_i^{\text{ref}}$ is fulfilled, i.e., $T=278~\text{K}$ in the initial phase. Because H$_2$O and N$_2$ are by far the most abundant components of the residual gas, an additional reference pair of partial-pressure and temperature values is defined to decrease the temperature gradient between the two temperature stages, so that, for instance, the maximal temperature becomes $200~\text{K}$ as long as $p_1$ is in the range between $5\times10^{-10}$ and $5\times10^{-9}~\text{mbar}$.

	\begin{table}[tbh]
	  \centering
	  \begin{tabular}{cccccc}
	    species (amu) & $T_{\text{adsor.}}$ [K] &\vline & $p^{\text{ref}}$ [mbar] & $T^{\text{ref}}$ [K] \\\hline
	    H$_2$O ($18$)  & 160 &\vline & $5\times10^{-9}$ & $278$\\
	     &  &\vline &   $5\times10^{-10}$ & $200$\\\hline
	    CO$_2$ ($44$)  & 125 &\vline & $5\times10^{-10}$ & $200$\\\hline
	    N$_2$ ($28$)  & 55 &\vline & $5\times10^{-9}$ & $150$\\
	     &  &\vline&    $5\times10^{-10}$ & $90$\\\hline
	    O$_2$ ($32$)  & 60 &\vline & $5\times10^{-10}$ & $90$\\
	  \end{tabular}
		\caption{Adsorption temperatures $T_{\text{adsor.}}$ and reference temperatures $T^{\text{ref}}$ and pressures $p^{\text{ref}}$ of residual gases in the vacuum chamber.}
	  \label{tab:setpoints}
\end{table}

When all partial pressures are below their reference values, the temperature feedback is stopped. The final temperature is then determined by the cooling power of the pulse-tube cooler, which corresponds to $3~\text{K}$ at the sample in our system. At this point, the partial pressures are below the detection limit of the RGA.

\begin{figure}[tb]
	  \centering
	  \includegraphics[width=86mm]{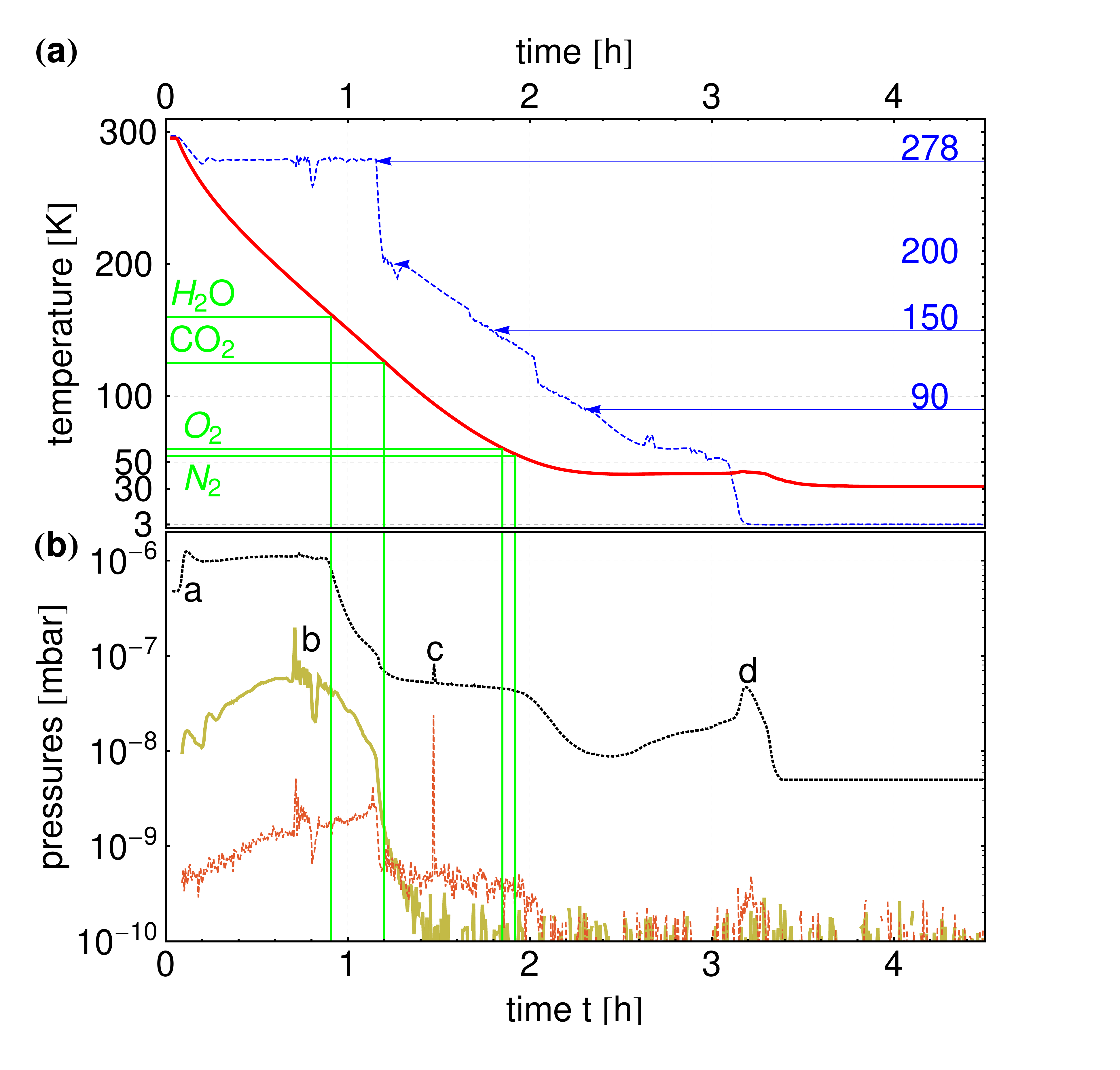}
	  \caption{(a) Temperatures of the sample holder (dashed blue) and $30~\text{K}$ stage (solid red) during the cooling procedure. The green lines indicate the temperatures at which critical gases freeze out. The blue lines show temperatures at which the sample holder is stabilized. (b) Typical pressures measured in the cryogenic chamber (dotted black) and the partial pressures of H$_2$O (solid yellow) and N$_2$ (dashed red) measured in the experimental chamber. The features at positions labeled a-d are discussed in the text.}
  \label{fig:cooldown}
	\end{figure}

Figure~\ref{fig:cooldown} illustrates this procedure by showing in panel (a) the temperatures of the sample holder (blue dashed trace) and the first stage (red solid trace) and in panel (b) the partial pressures of H$_2$O (yellow solid trace) and N$_2$ (red dashed trace) and the total pressure in the cryogenic chamber (black dotted trace).

The dark green lines indicate the position of the adsorption temperatures estimated from the sublimation points observed by an increase in partial pressure of the respective residual gas when letting the system warm up. The lower panel shows that the partial pressures of each residual gas decrease as soon as the adsorption point is reached. It also shows that, initially, the residual gas in the cryogenic chamber and the experimental chamber is dominated by water vapor.

We interpret four irregularities indicated by the labels a, b, c and d in the lower panel of Fig.~\ref{fig:cooldown} as follows:
Soon after the cooling procedure is started, the residual pressure in the cryogenic chamber rises (irregularity a) because the temperature of the high-temperature reservoir of the pulse-tube cooler (area shaded in red at the top of Fig.~\ref{fig:CoarseSetup}) rises. Irregularity b is an artefact of the feedback procedure because of a temporary failure caused by an inadequate time constant, which was then adapted manually. 

The sudden spike in the nitrogen partial pressure (irregularity c) is likely caused by a sudden release of trapped nitrogen gas (virtual leak). Finally, irregularity d is the result of the last rapid phase of cooling which leads to a temperature increase in a restricted area of the cryogenic chamber that exceeds the adsorption temperature of nitrogen for a short time. The fact that only the pressure in the cryogenic chamber rises, and not in the experimental chamber, demonstrates that gas does not diffuse from one chamber to the other. One sees from the temperature evolution of the sample holder that the main motivation to set the reference points ($p^{\text{ref}}$, $T^{\text{ref}}$) for CO$_2$ and O$_2$ in Table~\ref{tab:setpoints} is to slow down the cooling process and to prevent the residual gas from adsorbing at the sample holder.
The final temperatures of the first stage and the sample are $32.5~\text{K}$ and $3~\text{K}$, respectively. The corresponding total residual pressures are $<5\times 10^{-9}~\text{mbar}$ and $<10^{-9}~\text{mbar}$, respectively. The main and only measurable residual gas in the experimental chamber at the lowest temperatures is H$_2$ ($\approx6\times10^{-10}~\text{mbar}$).

\section{Measurement of the He$^*$ beam shape and position above the sample surface}	
\label{sec:distancecalibration}

\begin{figure}[!t]
\centering
	  \includegraphics[width=86mm]{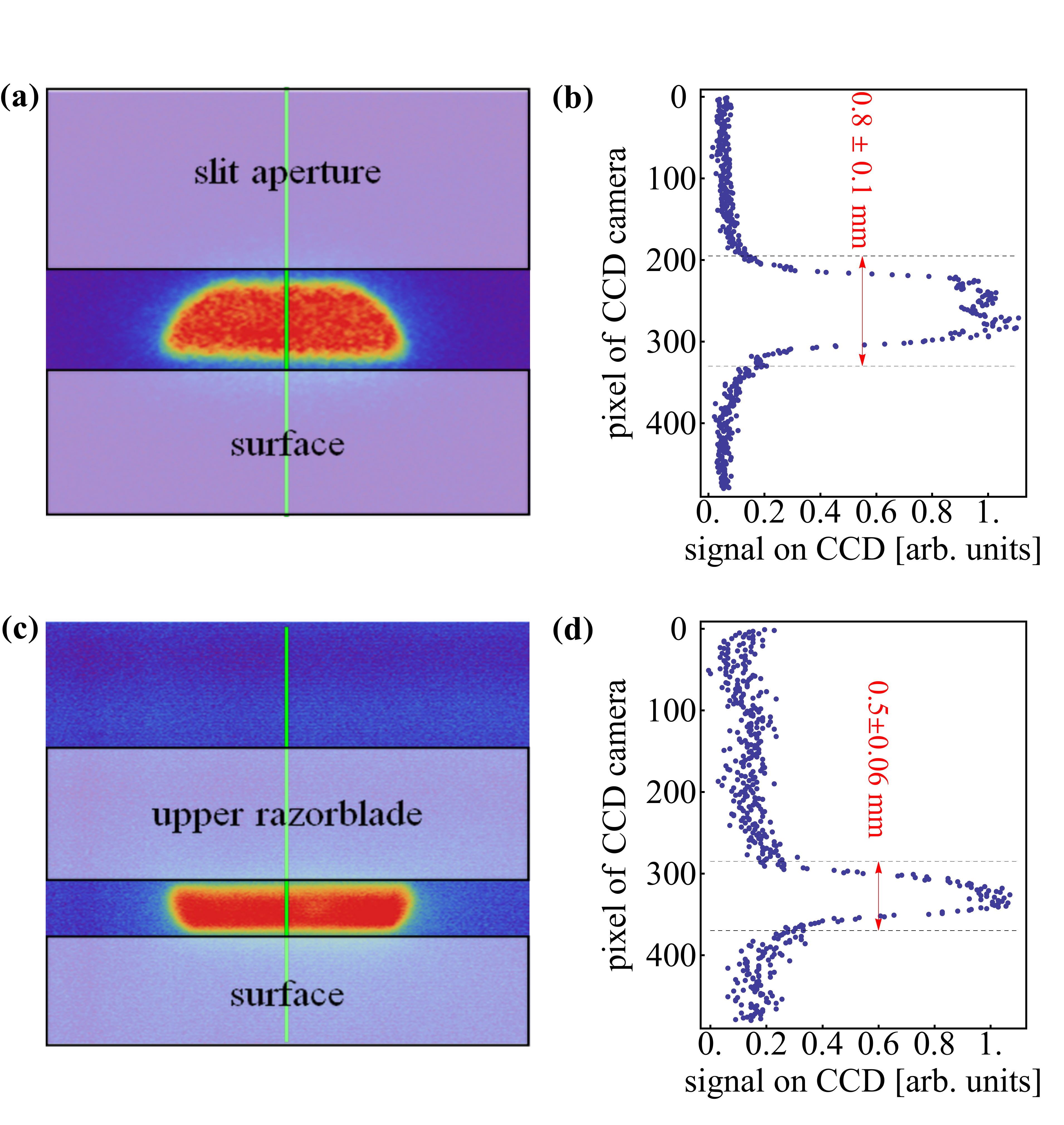}
	  \caption{Measurement of the position and width of the He$^*$ beam above the chip surface with an imaging MCP. Image obtained without (a) and with (c) collimating adjustable aperture. The intensity profiles along the vertical lines in panels (a) and (c) are displayed in panels (b) and (d), respectively (see text for details).}
	  \label{fig:AtomBeamImage}
\end{figure}

	To measure the size and position of the He$^*$ beam relative to the sample surface beyond the skimmer and the aperture, images of the beam as it collides with the MCP-phosphor-screen detector assembly are recorded with a CCD camera. The MCP front plate is located $16~\text{cm}$ downstream from electrode $5$ (see Fig.~\ref{fig:CoarseSetup}). To convert the image size and position measured at the detector into the size and position of the beam above the sample surface, a slit aperture of $0.8~\text{mm}$ width was placed immediately after the sample, with its lower edge at the sample surface. The beam image [see Fig.~\ref{fig:AtomBeamImage}(a)] was limited in its upper (lower) part by the upper (lower) edge of the aperture and on the sides by a circular hole of $6~\text{mm}$ diameter (not shown in Fig.~\ref{fig:CoarseSetup}) through which the beam passes when it exits the $3~\text{K}$ heat shield. The shape of the image and the knowledge of the geometric constraints provide a one-to-one correspondence between image pixels and small tubular volumes above the sample surface.

A pair of razor blades is then used to collimate the beam just before the photoexcitation region. The distance between the blades is reduced to typically $\approx500~\mu\text{m}$, i.e., less than the laser-beam width with the edge of each razor blade aligned parallel to the surface. The position of the blades are individually adjustable in the direction orthogonal to the surface so that they form an adjustable slit aperture. The cross-section of the beam, see Fig.~\ref{fig:AtomBeamImage}(c), is then constrained in its lower and upper parts by the razor blades. The absolute position of the beam center at $250~\mu\text{m}$ above the surface is derived from the image pixels and the previously determined one-to-one correspondence to the tubular volumes. The beam width (full width at half maximum) in the direction normal to the surface ($y$-direction) can be directly estimated from the detected intensity of the images along the yellow lines in Figures~\ref{fig:AtomBeamImage}(a) and (c). These are $0.8(1)~\text{mm}$ and $0.50(6)~\text{mm}$, respectively, in the presented example.

\newpage

\bibliographystyle{apsrev}

\end{document}